\documentclass[journal]{IEEEtai}

\usepackage{color,array}
\usepackage{graphicx}
\usepackage{amsmath,amssymb,amsfonts}
\usepackage{algorithmic,algorithm}
\usepackage{subfigure}
\usepackage{multirow}
\usepackage{enumerate}
\usepackage{orcidlink} 
\usepackage{booktabs}
\usepackage{url}
\usepackage{ulem}
\usepackage{soul}% Highlighting and strikeout
\usepackage{hyperref}
\hypersetup{colorlinks=true, linkcolor=blue, citecolor=blue, urlcolor=violet}
\begin{document}

\title{ISLS: IoT-Based Smart Lighting System for Improving Energy Conservation in Office Buildings}

\author{Peace Obioma\orcidlink{0009-0001-3020-6195}, Obinna Agbodike\orcidlink{0000-0003-1297-952X}, Jenhui Chen\orcidlink{0000-0002-8372-9221}, and Lei Wang\orcidlink{0000-0003-1810-3019} %

\thanks{This work was partly supported by the National Science and Technology Council, Taiwan, under Grant MOST-110-2221-E-182-041-MY3.} 
\thanks{P. Obioma is with the Department of Computer Science, University of Manchester, M139PL, England, (e-mail: obiomapeace2015@gmail.com).}
\thanks{O. Agbodike is with the Center for Artificial Intelligence in Medicine, Chang Gung Memorial Hospital, Guishan Dist, Taoyuan, 33375, Taiwan; and the \href{https://itrustal.com}{iTrustal.com}, (e-mail: obinnadyke@gmail.com).}
\thanks{J. Chen is with the Department of Computer Science and Information Engineering, Chang Gung University, Guishan Dist., Taoyuan 33302, Taiwan; the Center for Artificial Intelligence in Medicine, Chang Gung Memorial Hospital, Guishan Dist, Taoyuan, 33375, Taiwan, (e-mail: jhchen@mail.cgu.edu.tw).}
\thanks{L. Wang is with the Key Laboratory for Ubiquitous Network and Service Software of Liaoning Province, School of Software, Dalian University of Technology, China, (e-mail: lei.wang@dlut.edu.cn).}

}

\maketitle

\begin{abstract}

With the Internet of Things (IoT) fostering seamless device-to-human and device-to-device interactions, the domain of intelligent lighting systems have evolved beyond simple occupancy and daylight sensing towards autonomous monitoring and control of power consumption and illuminance levels. To this regard, this paper proposes a new do-it-yourself (DIY) IoT-based method of smart lighting system featuring an illuminance control algorithm. The design involves the integration of occupancy and presence sensors alongside a communication module, to enable real-time wireless interaction and remote monitoring of the system parameters from any location through an end-user application. A constrained optimization problem was formulated to determine the optimal dimming vector for achieving target illuminance at minimal power consumption. The simplex algorithm was used to solve this problem, and the system's performance was validated through both MATLAB simulations and real-world prototype testing in an indoor office environment. The obtained experimental results demonstrate substantial power savings across multiple user occupancy scenarios, achieving reductions of approx. 80\%, 48\%, and 26\% for one, two, and four user settings, respectively, in comparison to traditional basic lighting installation systems.

\end{abstract}


\section{Introduction}

\IEEEPARstart{A}{rtificial} lighting is an essential component of every building, serving both functional and aesthetic purposes. It not only enhances the visual appeal of indoor spaces but also provides the necessary illumination for a wide range of visual-based activities. Therefore, lighting systems play a vital role in ensuring environmental comfort and well-being.
However, while the need to provide illumination remains the primary focus of modern lighting systems, there is increasing emphasis on achieving this goal with maximum energy efficiency through technological advancement.
Unequivocally, artificial lighting contributes significantly to global energy consumption, representing nearly one-sixth of total electricity production worldwide. This underscores the urgent need for more energy-efficient systems \cite{Ziss'17}, especially with global sustainability goals in mind.
According to the International Energy Agency, transitioning to energy-efficient lighting systems could reduce global electricity consumption by over 1,500 terawatt-hours per annum, equivalent to the output of more than 600 medium-sized power plants \cite{IEA'20}. 

In (commercial office) buildings, artificial lighting accounts for over 25\% of electricity consumption and therefore represents a significant operational cost \cite{baharudin'21, angel'22}. Most traditional office lighting systems often lack adaptability to real-time occupancy changes and daylight availability, and coupled with typical employees' (i.e., users) negligence or indifference to switching off unused lights, further contributes to unnecessary energy wastage. Additionally, studies have shown that inefficient lighting in office environments not only impacts energy bills but also affects employee productivity and well-being \cite{mengya'24}. 

To reduce energy consumption rate while maintaining optimal indoor illuminance for occupants' visual comfort, automated smart lighting technologies have emerged as the viable and promising solution \cite{saffre'19, arun'17}. And as a fundamental component of future smart cities, the implementation of smart lighting systems not only offers significant energy conservation opportunities but also plays a crucial role in advancing global sustainability goals.
Moreover, with the recent advent of 5G and beyond (B5G) networks enabling internet of things (IoT) support for massive machine-type communications (MTC) \cite{Jenhui'20}, there exists a growing synergy between smart lighting systems and IoT. This synergy strengthens the case for IoT-based lighting solutions in office environments, where structured operating hours and predictable occupancy patterns make automated control systems particularly effective. The convergence of these dichotomous domains (i.e., lighting and IoT) is nowadays, driving the development of a broader ecosystem of MTC-enabled lighting devices, featuring controllable fixtures, dynamic illuminance adjustments, and auto-scheduling, all operable via application-based interfaces on remote networks.

In this paper, we present a low-cost, DIY IoT-based smart lighting system (ISLS) design, implemented with sensors and an illuminance control algorithm aimed at optimizing energy usage efficiency. The system provides dynamic, customizable indoor lighting adjustments to meet the users' specific photometric needs. The ISLS can autonomously adjust luminance levels and responds to changes in occupancy, availability of natural daylight, and other user-defined parameters. Thus, unlike existing smart lighting solutions, the proposed ISLS integrates both occupancy-based dimming and dynamic illuminance control, and in addition, its also provides real-time data analytics for tracking energy usage patterns for optimal energy management.

\section{Related Work}

By definition, a lighting system is considered smart when it can autonomously interact with external stimuli, enable connected luminaires to collaborate, and optimize energy consumption while aligning with user preferences \cite{Miki'04}. Recently, the advancements in smart lighting infrastructure have shifted toward data-driven analytical models with predictive control strategies \cite{Kumar'19}; enabling an optimal balance between natural daylight and artificial illumination to enhance energy-usage efficiency without compromising user experience \cite{Chraibi'18}.

\subsection{Smart Lighting in Urban Environments}

Prior research studies in the domain of intelligent lighting systems had predominantly focused on urban environments, where automated street lighting solutions pioneered the practical implementation of smart lighting. Notable examples include the intelligent street lighting system in \cite{He'19}, which integrates a wireless sensor network with a fuzzy decision-making algorithm to minimize energy consumption; the energy-efficient street lighting model in \cite{Didar'19}; and alternative approaches leveraging Long-Range (LoRA) technology to enhance connectivity and performance in smart street lighting systems \cite{Ezgi'19}; among others.

\subsection{Smart Lighting in Indoor Environments}

%However, research and development in smart lighting for indoor environments is rapidly progressing. 
With the rapid evolution of smart home and office automation, research into indoor smart lighting systems has gained traction.
In \cite{David'13}, a sensor-based networked lighting system was explored for automated illumination control, showcasing the potential of adaptive lighting adjustments. Similarly, \cite{Lee'19} introduced a context-aware intelligent lighting system capable of dynamically responding to environmental changes and user preferences. Whereas, other studies \cite{Milica'19} have demonstrated innovative approaches for energy-savings through advanced smart LED technologies for indoor lighting purposes.

\subsection{IoT-Driven Advancements}

More recently, the integration of IoT in smart lighting is the research interest in-vogue since IoT became the integral component of modern smart homes and cities. In \cite{Lee'22}, a `leader-follower' smart office lighting approach was introduced, and likewise \cite{Chun'22} proposed a `master-slave' intelligent LED control system; both demonstrating how IoT-based distributed control enhances energy efficiency and system reliability. Additionally, an IoT-based power conservation model was developed in \cite{Theddu'22} to minimize unnecessary energy consumption, and in \cite{Hoss'17}, an intelligent light-dimming mechanism using support vector machines is proposed, underscoring the expanding role of machine learning in smart lighting systems.

\subsection{Addressing the Gap in Literature}

Despite the above-mentioned advancements, the need for a unified solution that optimally balances energy efficiency, real-time monitoring, and user adaptability within a cost-effective framework persists. Our proposed IoT-based Smart Lighting System (ISLS) addresses these challenges by utilizing real-time sensor data for automated luminance adjustments, enabling remote control, and generating actionable insights through data-driven performance monitoring. Through both MATLAB simulations and a low-cost, custom DIY IoT prototype implementation, the ISLS proves to not only enhance energy efficiency by dynamically adjusting lighting usage patterns, but also to be viable in fostering the development of sustainable, intelligent lighting system that is adaptive to user preferences and environmental conditions; thus ultimately making smart lighting more accessible, scalable, and cost-effective.

\section{SYSTEM MODEL}

Many previous methodologies aimed at optimizing energy efficiency and occupant satisfaction in lighting systems often encounter the fundamental challenge of light usage control. Recognizing this as a critical bottleneck, we classify control systems into three main categories: the logic-based, regulation-based, and optimization-based controllers. 

\begin{itemize}
    \item Logic-based method uses varying decision-making techniques (i.e., logic circuits) to determine appropriate lighting actions based on sensor data.
    \item Regulation-based method maintains closed-loop stability by ensuring that the illumination levels track a predefined reference or set-point value.
    \item Optimization-based method formulate lighting control decision as an optimization problem, leveraging mathematical techniques to enhance energy efficiency.
\end{itemize}

Each method offers unique advantages, but our work specifically focuses on optimization-based control to achieve a balanced energy efficient lighting system implemented with LED luminaires.

\begin{figure}
\centering
\includegraphics[width=15pc]{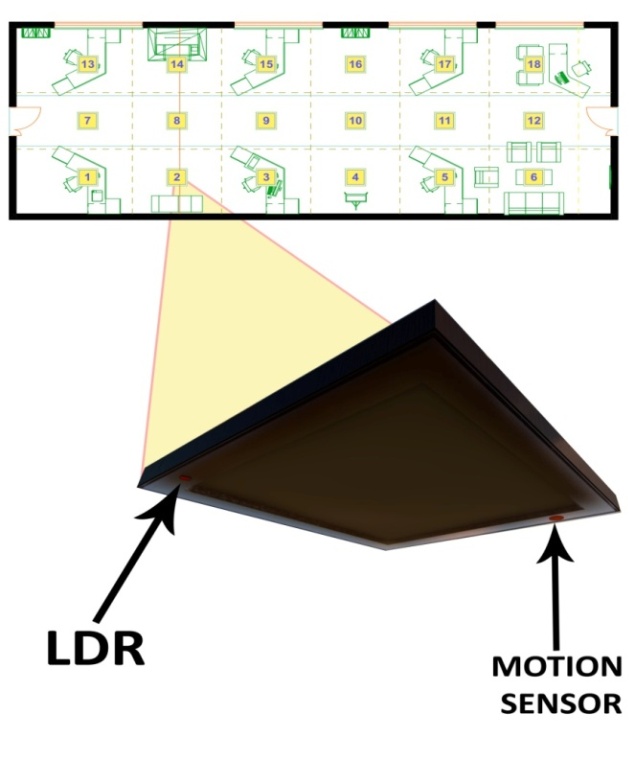}
\caption{Floor plan illustrating an indoor space projecting the luminaire}\label{floorplan}
\end{figure}

In this study, we evaluated the lighting system in a standard workplace environment. The workspace consists of two planes: the bottom plane, divided into $N$ logical zones, and the top plane, which houses multiple ceiling-mounted luminaires equipped with light sensors. A typical indoor office illumination floor plan is illustrated in Fig.~\ref{floorplan}

We explored a distributed luminaire optimization-based control system whereby each luminaire is integrated with an LED, a passive infrared (PIR) sensor, light sensor, LED driver, and the controller module. The PIR sensor detects local occupancy or unoccupancy within a designated logical zone; while the light sensor - a light-dependent resistor (LDR), measures the net light levels in the corresponding area.

\begin{figure}
\centering
\includegraphics[width=20pc]{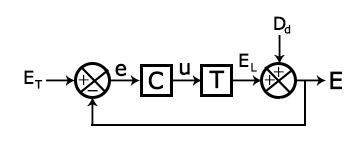}
\caption{Block diagram illustration of the closed loop system.}\label{closeloop}
\end{figure}

The choice of LED luminaire used in the ISLS prototype were determined by factors such as low cost, durability, and brightness. In addition, LEDs allow the design to be modeled as a linear system \cite{Marco'15, Sepehr'14}. Adopting a close-loop control approach for the prototype, we characterize the input-output relationship of the lighting system as depicted in Fig.~\ref{closeloop}. The variables' denotation are defined as follows: $E_T$ = target illuminance value, $e$ = error term, $C$ = control algorithm, $u$ = control variable given by ${[u_i]}^T \in \mathbf{R}$, $T$ = light transpose matrix, $E_L$ = light fixture illuminance, $D_d$ = daylight disturbance, and $E$ = actual values of the illuminance.

Therein also, the $T$ function decides the input/output relationship of the system, given that the LEDs are connected linearly. To model the lighting system, we viewed the light fixtures as subsystems operated by LED drivers which give a specific luminous flux output. Thus, $T$ portrays the lighting system as having a linear connection between the output and the input of the system. According to \cite{Hui'09} the total luminous flux of an LED system with $N$ LED devices is directly linked to the power of the LED, given by
\begin{equation}\label{eq1}
\varphi = N\eta{P_d},
\end{equation}
where $\varphi$ is the total luminous flux, $\eta$ denotes the luminous efficacy, and ${P_d}$ represents the real power of one LED (in watt). For in-depth definition, the expression in (\ref{eq1}) can be further expanded as,
\begin{equation}\label{eq2}
\varphi = N{\eta_o}[1 + {K_e}(T_{\alpha} - {T_o})P_d + {K_e}{K_h}(R_{jc} + NR_{hs}){P_d}^2],
\end{equation}
where $\eta_o$ denotes the LED rated efficacy, $T_o$ is the rated temperature, $R_{jc}$ is the thermal resistance of the LED, $R_{hs}$ is the heat-sink thermal resistance, $K_e$ rep the relative rate of reduction of efficacy with increasing temperature, and $K_h$ is the heat dissipated from the LEDs. In this regard, the luminous flux output of the $ith$ LED can be expressed as
\begin{equation}\label{eq3}
\varphi_i = \alpha_i {P_d},
\end{equation}
where $\alpha_i$ is the coefficient which relates the luminous flux of the $ith$ LED to the input power. It is further shown in \cite{Hui'09} that the output luminous flux $\varphi$ could be related to its photometric, electrical, and thermal properties. Hence, (\ref{eq2}) is reformulated as
\begin{equation}\label{eq4}
\varphi = n{E_o}[1 + {K_e}(t_{\alpha} - {t_o})P_d + {K_e}{K_h}(R_{jc} + NR_{hs}){P_d}^2].
\end{equation}
If we consider a lighting system with $n$ point light sources, and $m$ light sensors, the relationship between the illuminance and luminous flux can be represented using the inverse square law \cite{Sepehr'14}
\begin{equation}\label{eq5}
\mathbf{E} = \frac{1}{4\pi}\sum_{i=1}^{n}\frac{\varphi_n}{D_{mn}^2},
\end{equation}
where $\mathbf{E}$ is an  $(m \times 1)$ vector representing the illuminance measured by the $mth$ sensor, $\varphi_n$ is the $(n \times 1)$ vector representing the luminous flux and $D_{mn}$ is a $(m \times n)$ matrix indicating the distance between the $mth$ light sensor and $nth$ light source. By Expanding (\ref{eq5}) we have
\begin{equation}\label{eq6}
\unskip
%\begin{align}
\begin{bmatrix} E_1 \\ E_2 \\ \vdots \\ E_m \end{bmatrix} = \frac{1}{4\pi}
\begin{bmatrix} \frac{1}{D_{11}^2} & \frac{1}{D_{12}^2} & \cdots & \frac{1}{D_{1n}^2}\\
\frac{1}{D_{21}^2} & \frac{1}{D_{22}^2} & \cdots & \frac{1}{D_{2n}^2}\\
\vdots & \vdots & \vdots & \vdots \\
\frac{1}{D_{21}^2} & \frac{1}{D_{22}^2} & \cdots & \frac{1}{D_{2n}^2}\\ \end{bmatrix}
\begin{bmatrix} \varphi_1 \\ \varphi_2 \\ \vdots \\ \varphi_n \end{bmatrix}.
%\end{align}
\end{equation}

\noindent Applying (\ref{eq3}) to (\ref{eq6}), we have an extended matrix expressed as %(to_verify_app_eq2 or eq3)%
\begin{equation}\label{eq7}
\unskip
%\begin{align}
\begin{bmatrix} E_1 \\ E_2 \\ \vdots \\ E_m \end{bmatrix} = \frac{1}{4\pi}
\begin{bmatrix} \frac{1}{D_{11}^2} & \frac{1}{D_{12}^2} & \cdots & \frac{1}{D_{1n}^2}\\
\frac{1}{D_{21}^2} & \frac{1}{D_{22}^2} & \cdots & \frac{1}{D_{2n}^2}\\
\vdots & \vdots & \vdots & \vdots \\
\frac{1}{D_{21}^2} & \frac{1}{D_{22}^2} & \cdots & \frac{1}{D_{2n}^2}\\ \end{bmatrix}
\begin{bmatrix} \varphi_1{P_{d1}} \\ \varphi_2{P_{d2}} \\ \vdots \\ \varphi_n{P_{dn}} \end{bmatrix}.
%\end{align}
\end{equation}

In (\ref{eq7}), the linearity assumption of the lighting system model indicates a good estimation of the system as it shows that the output illuminance is linearly related to the input power of the LEDs. Recall that $T$ determines the relationship between the input and output of the system. This can be estimated by posing the model identity problem as a least square problem \cite{Imam'17} which can be costly. Meanwhile, to ensure minimal model complexity, we considered the first two luminaires (denoted as $LUM1$ and $LUM2$) together with their respective photo-sensors (denoted as $\rho1$ and $\rho2$). Then, $T$ is modeled as a $2\times 2$ matrix expressed as
\begin{equation}\label{eq8}
\unskip
T_{nm} = \begin{bmatrix} T_{11} & T_{12} \\ T_{21} & T_{22} \end{bmatrix},
\end{equation}
where $T_{nm}$ represents the relationship between the $nth$ luminaire and the $mth$ photo-sensor. In this work, the sensors are coupled to the luminaires, therefore we surmise $m$ = $n$.

Two respective experiments were conducted to investigate the luminaire identification process. Firstly, we fed different dimming levels to $LUM1$ while $LUM2$ was turned OFF, and then observed values from $\rho1$ and $\rho2$ were noted. Secondly, $LUM1$ was turned OFF while the dimming levels were applied to $LUM2$ during which the values from $\rho1$ and $\rho2$ were obtained. In each of these experiments, only one luminaire was turned on to ensure that the effect of disturbance (i.e., illumination interference) from neighboring light fixtures would not be recorded by the photo-sensors. A multiple linear regression approach \cite{Stroup'08} was used to formulate the system model response, given by
\begin{equation}\label{eq9}
Y_i = \beta_0 + \beta_1 X_{i1} + \beta_2 X_{i2} + \cdots + \beta_k X_{ik} + \epsilon_i \forall i = 1,2,\cdots,n.
\end{equation}

\noindent Translating (\ref{eq9}) in matrix form, we have
\begin{equation}\label{eq10}
Y = X\beta + \epsilon,
\end{equation}
where,
\begin{equation}\label{eq11}
\unskip
Y = \begin{bmatrix} Y_1 \\ \vdots \\ Y_n \end{bmatrix},
X = \begin{bmatrix} 1 & X_{11} & \cdots & X_{1k} \\  1 & X_{21} & \cdots & X_{2k} \\ \vdots & \vdots & \cdots & \vdots \\ 1 & X_{n1} & \cdots & X_{nk} \end{bmatrix},
\end{equation}
\begin{equation}
\beta = \begin{bmatrix} \beta_0 \\ \vdots \\ \beta_k \end{bmatrix},\ \text{and}\ \epsilon = \begin{bmatrix} \epsilon_0 \\ \vdots \\ \epsilon_n \end{bmatrix}
\end{equation}
given that $y$ is a $(n \times 1)$ vector of the observations, $X$ is an $(x \times p)$ matrix of the levels of the independent variables, $\beta$ is a $(p \times 1)$ vector of the regression coefficients, and $\epsilon$ is a $(n \times 1)$ vector of the random errors.

To find the vector of the least square estimators $\hat{\beta}$ that minimizes, we have that
\begin{equation}\label{eq12}
\hat{\beta} = \sum_{i=1}^{n}\epsilon_{i^2} = \epsilon^{T} \epsilon = (y - X\beta)^T(y - X\beta).
\end{equation}

Thus, $\hat{\beta}$ estimator is the solution for $\beta$ in finding the derivative, given that $\frac{\partial \hat{\beta}}{\partial \beta} = 0$,

Then, the derivative can be resolved to
\begin{equation}\label{eq13}
X^T X \hat{\beta} = X^T Y.
\end{equation}

Thereafter, $\beta$ can be estimated by reformulating (\ref{eq12}) as
\begin{equation}\label{eq14}
\hat{\beta} = (X^T X)^-1 X^T Y.
\end{equation}

Using MATLAB codes to solve $(\ref{eq14})$, we use the values of $LUM1$ and $LUM2$ obtained from the two respective experiments to establish that
\begin{equation}\label{eq15}
Y_1 = 0.02 X_1 + 0.21 X_2 + 4.76
\end{equation}
and
\begin{equation}\label{eq16}
Y_2 = 0.018 X_1 + 0.02 X_2 + 2.53.
\end{equation}

Expressing $(\ref{eq15})$ and $(\ref{eq16})$ in matrix form, we have
\begin{equation}\label{eq17}
\unskip
\begin{bmatrix} Y_1 \\ Y_2 \end{bmatrix} = \begin{bmatrix} T_{11} & T_{12} \\ T_{21} & T_{22} \end{bmatrix} + \begin{bmatrix} B_1 \\ B_2 \end{bmatrix}.
\end{equation}
Substituting variables in (\ref{eq17}) with the values obtained from MATLAB, we have
\begin{equation}\label{eq18}
\unskip
\begin{bmatrix} Y_1 \\ Y_2 \end{bmatrix} = \begin{bmatrix} 0.02 & 0.21 \\ 0.18 & 0.02 \end{bmatrix} + \begin{bmatrix} 4.76 \\ 2.53 \end{bmatrix}.
\end{equation}

The offset matrix $B$ in (\ref{eq17}) whose values are given in (\ref{eq18}) represents the systemic uncertainty which is also handled by the system. Note that to attain non-perceptible flicker, Pulse Width Modulation (PWM) was applied to regulate the power consumption of the luminaire while delivering the target illuminance. Whereas, the dimming level is a function of the duty cycle of the waveform whose amplitude is the illuminance level and is determined by the radiation pattern of the LEDs \cite{Yang'10}. For $N$ logical zones in the indoor workspace, with $n$ number of luminaires (see Fig.~\ref{floorplan}), let $d$ be the dimming vector with dimming levels $d = \{d_1, d_2, \cdots, d_n\}$. Hence $d_N$ being dimming level of $nth$ luminaire implies $d_N \in \mathbf{R}$ and $0 \leq d_n \leq 1$; where $d_N = 0$ represents an LED completely dimmed to OFF state and $d_N = 1$ represents maximum illuminance. Furthermore, the average power consumed by an LED at any given instance can be given by,
\begin{equation}\label{eq19}
P_x d_N,
\end{equation}
where $x$ represents either the ON or OFF state. Similar to \cite{Didar'19}, our model presents the lighting system as a multi-point light source consisting of spatially distributed LEDs. To achieve uniform illumination, we employed an array of LEDs distributed at the top plane (i.e., ceiling) of the workspace. Assuming that the respective LEDs has a characteristic emitting pattern which can be considered as a function of the solid angle $f(\theta)$ in a 3-Dimensional indoor space; thus mapping $f(\theta)$ into the workspace plane at a height distance $h$ from the ceiling we would formulate 
\begin{equation}\label{eq20}
f(x,y;h) = \frac{f(\theta)}{x^2 + y^2 + h^2},
\end{equation}
where $f(x,y:h)$ is the centric illuminating pattern, while $x$ and $y$ represents the in-space coordinate of the lighting point. Dimming the LEDs results to a dimmed illumination pattern. With respect to (\ref{eq20}), we aim to achieve a uniform distribution of illumination across the room for occupied zones and minimal illumination for the unoccupied zones. Let us denote the target illuminance levels as $E_T^o$ and $E_T^u$ for occupied and unoccupied zones, respectively. The illuminance from the $nth$ luminaire at point $(x,y)$ and height $h$ with the absence of daylight given that other luminaires are turned off, can be expressed as $E_n (x,y:h)$. At dimming level $d_n$, the illuminance becomes $d_n E_n(y,y:h)$. For daylight, we consider daylight disturbance as $D_d(x,y:h)$, thus total illuminance of the system can be mathematically given as,
\begin{equation}\label{eq21}
E_T (x,y;h;d) = \sum_{n=1}^{N} d_N E_n (x,y;h) + D_d(x,y;h).
\end{equation}

The distortion in illumination pattern at location point (x, y) with respect to a target illuminance level is characterized by the illumination contrast. It is expected that variations in the illumination level as given by the dimming vector $d$ with respect to $E_T^0$ must be below the illumination contrast threshold denoted as $C_t$. This was presented by Weber's law \cite{Caicedo'10}
\begin{equation}\label{eq22}
C(E_T (x,y;h;d),E_T^0  = \frac{D E_n(x,y;h) - E_T}{E_T} \leq C_t.
\end{equation}

With respect to (\ref{eq19}), a constrained optimization problem is formulated to determine the optimum dimming vector that would yield minimal power consumption; given by
\begin{equation}\label{eq23}
\begin{split}
\hat{D} = \arg\min{D} \sum_{n=1}^{N}P_x D_N \\
s.t \left\{ \,
\begin{IEEEeqnarraybox}[][c]{l?s}
\IEEEstrut
E (x,y;h) + D_d \geq E_{Tn} \\
0 \leq D_N \leq, N = 1, \cdots, N \\
D_N E_n(x,y;h) - E_T \leq C_t E_T
\IEEEstrut
\end{IEEEeqnarraybox}
\right.
%{E (x,y;h) + D_d \geq E_{Tn}}{0 \leq D_N \leq, N = 1, \cdots, N}\\
%{D_n E_n(x,y;h) - E_T \leq C_t E_T}
\end{split}
\end{equation}

\noindent given that $P_{x}D_{N}$ is the power consumed by luminaire  at dimming level $d_N$, $E(x,y;h)$ is the illuminance at a given logical zone in the workspace, $D_d$ is the daylight distribution, and $C_t$ is the allowable illumination contrast threshold.  

\noindent NB: the objective function and constraints in (\ref{eq23}) are linear and thus considered a linear programming problem.

\begin{figure*}[th]
\centering
\includegraphics[width=32pc]{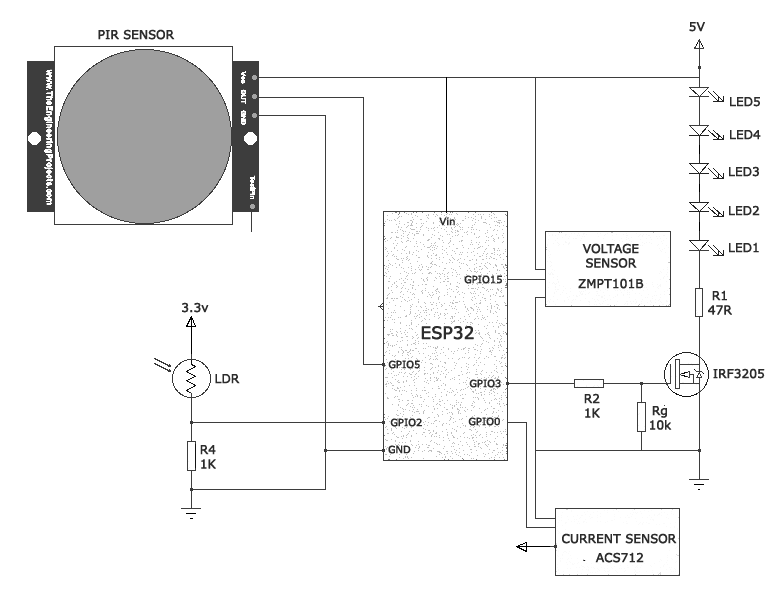}
\caption{The circuit diagram of the IoT system prototype.}\label{circuit}
\end{figure*}

%\section{SYSTEM ARCHITECTURE}
\section{DESIGN COMPONENTS}\label{components}

Overall, the proposed ISLS is divided into three functional parts: the cloud server, the user application and the IoT-based hardware.

\begin{figure}
\centering
\includegraphics[width=17pc]{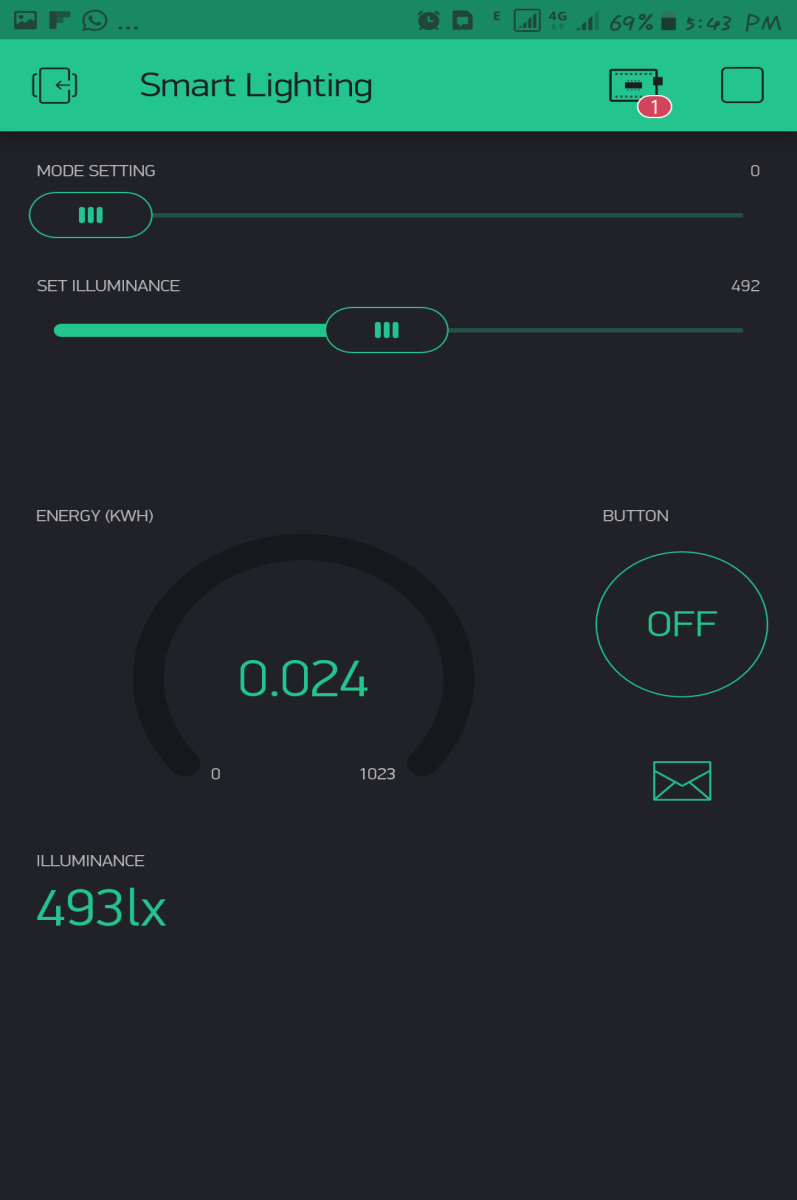}
\caption{End-user luminance control application (widget) interface.}\label{widget}
\end{figure}

\begin{enumerate}[I.]

\item \textbf{The cloud server} We adopted the Blynk server \cite{Blynk'19} for the remote monitoring and control of our proposed ISLS. It is an open-source Netty based Java server responsible for transmitting communications between a mobile application and several microcontrollers. \cite{Blynk'19} allows information hosting on a local server and supplies the localhost with built-in SSL certificates for security, and provides administrative interface accessible via a URL login process with dedicated IP address and port number.
\item \textbf{The user application} We designed a widget interface (see Fig.~\ref{widget}) to enable end-user interaction with the system over the network (on native mobile devices such as IOS or Android operating systems). For this task, the Blynk application library was installed on Arduino IDE, after which the desired microcontroller board was selected and the program code for the system was written and embedded into the microcontroller implemented with an authentication token associated to each controlled luminaire. The widgets used for the user interface are as follows;
    \begin{enumerate}
    \item Slider widget: labeled as “SET ILLUMINANCE” is used for set the illuminance to any value desired by the user
    \item Gauge widget: labeled as “ILLUMINANCE” is used as a feedback mechanism to show the illuminance status of the environment. When a user sets a desired illuminance in system, this widget is responsible for providing feedback on the illuminance using the in-built sensors.
    \item Button widget: is used to switch the lighting system on or off.
    \item Mode slider widget: is a real-time scheduling functionality which enables the lighting system to dynamically adjust according to certain day and night modes regardless of sensor readings. For example, in this task, we set from 9:00 to 18:00 hours as the time duration whereby the fixtures are automatically turned off (based on the assumption that from 9am, sufficient natural daylight has entered the indoor space until 6pm). The night mode operation commences after 6pm. Thresholds are set in the code to prevent fluctuating of the luminaries whenever a change in sensor readings occurs. Furthermore, the control algorithm ensures that variations must exceed certain thresholds before actions are initiated. Other features include automatic switch-off of the luminaire (after 5 minutes delay) during day time if no activity is detected in unoccupied zones of the indoor space.
    \item Email widget: enables users to receive an email comprising a concise summary of energy usage and cost on a weekly basis.
    \item Power gauge widget: displays the power consumed in real time.
    \item Device selector widget: facilitates the authentication token for effective monitoring and control of several luminaires from the user application.
    \end{enumerate}
\item \textbf{The IoT prototype} This consists of the microcontroller, PIR sensor, ambient sensor, voltage sensor, and current sensor. We used the ESP32 microcontroller \cite{ESP32} because it is a low powered integrated circuit with on-chip Wi-Fi shield and Bluetooth module, respectively, and also supports PWM dimming, and analog to digital (ADC) conversions. The other sensors are resource-constrained devices, for sensing and monitoring of physical environmental conditions \cite{Blynk'19}. The ZMPT101B voltage sensor and ACS712 current sensor were used in this work to acquire information on the power and energy consumed, and the luminaires used in the design is ``Cree XLAMP MX-6 high brightness LEDs'' with an output of up to 114 lumens. The complete DIY IoT circuit design is shown in Fig.~\ref{circuit}, and the cost breakdown of each hardware component is outlined in Table~\ref{tab:costs}. %, highlighting the system's simplicity and affordability.
    \end{enumerate}

\begin{table}[htb]
\centering
\caption{Cost breakdown of the prototype components}\label{tab:costs}
\begin{tabular}{l c}
\hline \vspace{0.1cm}
\textbf{ITEMS}                        & \hspace{2cm} \textbf{\$(USD)}  \\ \hline
ESP32 NodeMCU                         & \hspace{2cm} 1.00          	\\ [1.5mm]
LDR sensor                            & \hspace{2cm} 0.50         	\\ [1.5mm]
ZMPT101B Voltage sensor               & \hspace{2cm} 0.42        	\\ [1.5mm]
ACS712 Current sensor                 & \hspace{2cm} 0.68        	\\ [1.5mm]
IRF3205 Mosfet                        & \hspace{2cm} 1.43        	\\ [1.5mm]
LEDs (5 pieces)                       & \hspace{2cm} 0.50         	\\ [1.5mm]
Resistors (4 pieces)                  & \hspace{2cm} 0.10         	\\ [1.5mm]
Printed Circuit Board (PCB)           & \hspace{2cm} 4.00           \\ [1.5mm]
Casing (circuit enclosure)            & \hspace{2cm} 6.00           \\ [1.5mm]
Solder Lead wire                      & \hspace{2cm} 0.50         	\\ [1.5mm]\hline
\textbf{Total cost}                   & \hspace{2cm} \textbf{\$15.13} 	\\ \hline
\end{tabular}
\end{table} 

The system's design components and their corresponding prices in Table~\ref{tab:costs} were sourced from an online store\footnote{AliExpress: \url{https://aliexpress.com}}, and are valid as at the time of this publication. With a total cost of \$15.13 per luminaire unit, the ISLS prototype demonstrates the system's economic viability for wide-scale deployment in office buildings. Compared to commercial smart lighting solutions that typically cost \$50-100 per unit, our DIY approach achieves competent functionality at approximately 70-85\% lower cost. This cost-effectiveness is particularly relevant for retrofitting existing office buildings, where the return on investment period for lighting upgrades is a crucial consideration. And moreover, the low-cost nature of the system does not compromise its functionality, as demonstrated by the performance results undermentioned in Section~\ref{deploy}.

\section{IMPLEMENTATION}\label{deploy}

Here, we present the resulting outcomes of our simulation and real-world experiments, validating the ISLS performance. For the simulation experiment, we assumed the values of the experimental parameters (see Table~\ref{tab:params}) which are consistent with basic workspace lighting norm. The simplex algorithm\footnote{Simplex: \url{https://github.com/LKolonias/Simplex-Method}} was used to obtain the dimming levels of the LED luminaires, via solving the linear programming problem formulated in (\ref{eq23}) to find the optimal objective function (see Algorithm~\ref{algo1}). The objective function seeks to minimize power consumption while satisfying illuminance constraints through iterative optimization of the basic feasible solution. The algorithm particularly suits our problem as it efficiently handles the bounded variables ($0 \leq D_N \leq 1$) and linear illuminance constraints.

\begin{algorithm}
\caption{Optimal Illuminance Dimming for ISLS}\label{algo1}
\begin{algorithmic}[1]
\REQUIRE Power consumption rates $P_x$, luminaires $N$, illuminance threshold $E_T$, environmental factors $E(x, y, h)$, and constraints
\ENSURE Optimal dimming vector $d^*$

\STATE Initialize dimming values $d_N$ for each luminaire $N$
\STATE Define objective: Minimize $\sum (P_x \cdot d_N)$

\FOR{each luminaire $N$}
    \STATE Ensure illuminance meets threshold: $E(x, y, h) + D_d \geq E_T[N]$
    \STATE Constrain dimming levels: $0 \leq d_N \leq 1$
    \STATE Ensure compliance: $d_N \cdot E_N(x, y, h) - E_T \leq C_t \cdot E_T$
\ENDFOR

\STATE Solve using Simplex algorithm (MATLAB): 
\STATE $d^* = \text{linprog}(\text{objective, constraints})$
%\STATE $d^* = \text{simplex\_solver}(\text{objective, constraints})$

\RETURN Optimal dimming vector $d^*$
\end{algorithmic}
\end{algorithm}

\begin{table}
\centering
\caption{Standard parameter values for MATLAB simulation}\label{tab:params}
\begin{tabular}{cc}
\hline
\multicolumn{1}{c}{\textbf{Parameters}} & \multicolumn{1}{c}{\hspace{2cm}\textbf{Values}} \\ [1.5mm]
\hline\hline
$E_T^o$ & \hspace{2cm} 600Lux \\ [1.5mm]
$E_T^u$ & \hspace{2cm} 300Lux  \\ [1.5mm]
$C_t$   & \hspace{2cm} 0.3      \\ [1.5mm]
$N$     & \hspace{2cm} 4         \\ [1.5mm]
\hline
\vspace{0.1cm}
\end{tabular}
\end{table}

\begin{figure}[ht!]
        \centering
            \subfigure[Time series plot of LUM1]
            {
                %\label{Time series plot of $lambda1$}
                \includegraphics[width=.22\textwidth]{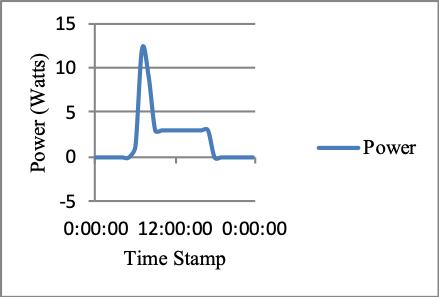} % .png .jpg ... according to supported graphics files
            }
            \subfigure[Time series plot of LUM2]
            {
                %\label{Time series plot of $lambda2$}
                \includegraphics[width=.22\textwidth]{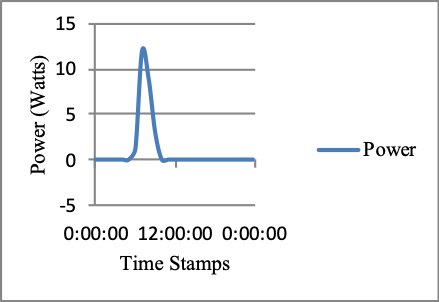} % .png .jpg ... according to supported graphics files
            }\\
            \subfigure[Time series plot of LUM3]
            {
                %\label{Time series plot of $lambda3$}
                \includegraphics[width=.22\textwidth]{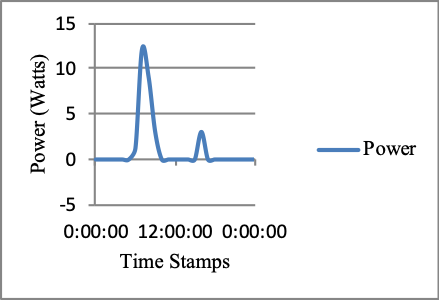} % .png .jpg ... according to supported graphics files
            } % for new row or line of subfigures
            \subfigure[Time series plot of LUM4]
            {
                %\label{Time series plot of $lambda4$}
                \includegraphics[width=.22\textwidth]{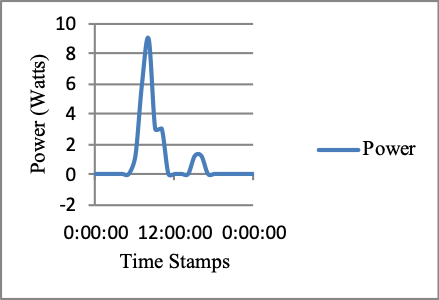} % .png .jpg ... according to supported graphics files
            }

        \caption{Plot showing power consumption rates in scenarios (a), (b), (c) and (d) for $LUM1,\ldots,LUM4$, respectively (obtained from the sensor values), during a 24hr time-series activity in the workspace.} 
        \label{timeseries}
        \end{figure}

Given for instance, a $15 \times 9$ft indoor workspace environment (with 2 windows allowing day-time natural light) equipped with existing light fixtures -- four (4) wired 12-watt LED-based luminaries referred to as the base installation, we integrate the proposed DIY IoT prototype comprising the components described in Section~\ref{components}. %%the wireless communication module, and occupancy/photo sensors to the base installation (i.e., existing light fixtures).% 
Without user interference via the widget application, we obtained the initial light usage data collected by the sensors in the system over a 24-hour period and observed the time-series activity of the four luminaries - to ascertain the amount of power consumed by the luminaires in the ISLS system. A graph plot of the energy consumption rates are shown in Fig.~\ref{timeseries}. Note that $LUM2$ and $LUM3$ luminaires are located closer to the windows, hence due to natural light interference (triggering the ISLS auto-illuminance dimming) they consume lesser power than $LUM1$ and $LUM4$ during the time of operation and occupancy. Overall, the energy consumption of the luminaires is calculated as,
\begin{equation}\label{eq25}
\sum_{n=1}^{N} (P \times d)\gamma
\end{equation}

where $\gamma$ is the number of active hours (summarized as the time stamps), and $d$ is the dimming value. 

Meanwhile, recall that the light fixtures of the base installation is not dimmable, therefore the luminaires continuously works at maximum output, hence $d$ = 1 (without the ISLS deployment). So for example, given a 12-watt LED luminaire, the total energy(KW/hr) over a 12 hour (active) period = $\frac{12\times 1 \times 12}{1000}$ = 0.144 KW/hr per luminaire. This calculation aligns with standard methods for determining energy usage, where energy (in kilowatt-hours) is the product of power (in kilowatts) and time (in hours). Therefore, considering 4 luminaires ($LUM1, \cdots, LUM4$) used in this experiment, $0.144 \times 4 = 0.576$ KW/hr total energy usage. %% 

\begin{figure}
\centering
\includegraphics[width=20pc]{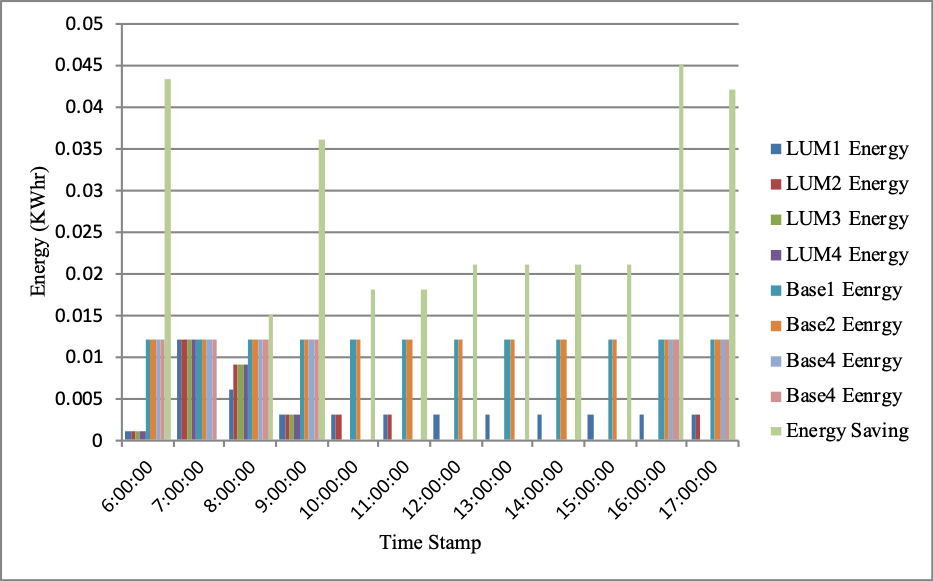}\\
\caption{Energy profile for one occupant scenario}\label{profile1}
%\end{figure}
\vspace{2mm}
%\begin{figure}
%\centering
\includegraphics[width=20pc]{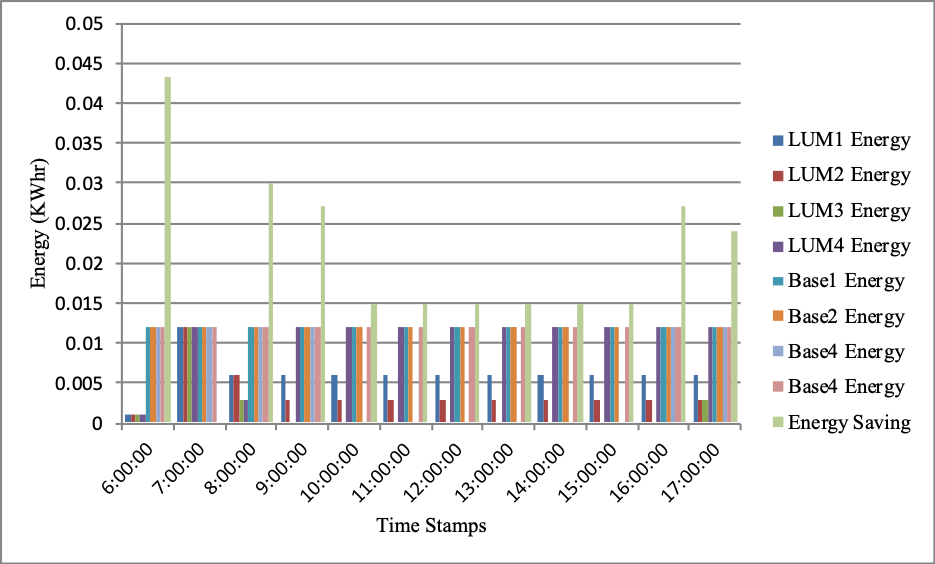}
\caption{Energy profile for two occupants scenario}\label{profile2}
\vspace{2mm}
%\begin{figure}
%\centering
\includegraphics[width=20pc]{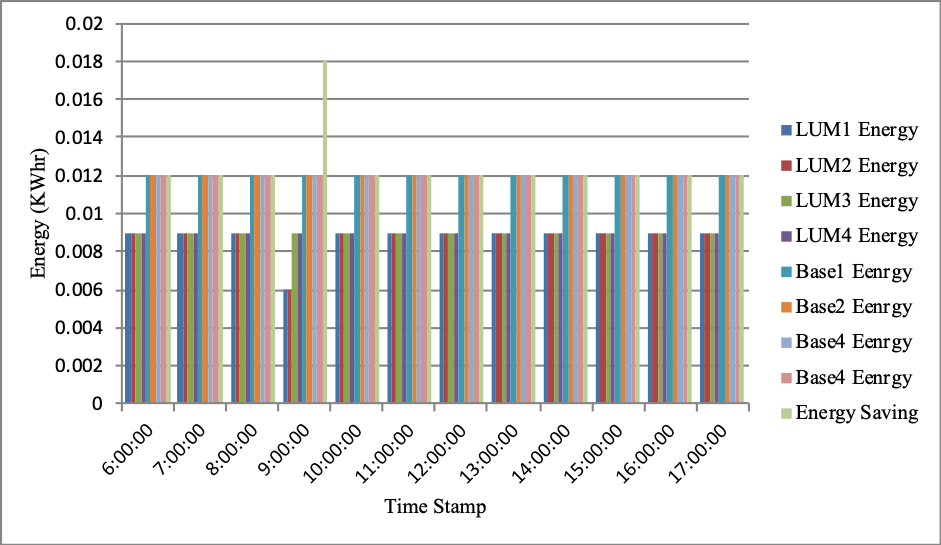}
\caption{Energy profile for four occupants scenario}\label{profile4}
\end{figure}

\subsection{Discussion and Analysis}
%%\section{Discussion}
Based on the numeric data values from the sensor readings during a time period of 6:00 to 17:00 (typical office hours), we first compared the daytime energy consumption rate of the base installation against that of the proposed ISLS. The resulting outcomes were plotted in Fig.~\ref{profile1}, Fig.~\ref{profile2} and Fig.~\ref{profile4} scenarios comprising of a one, two, and four occupants, respectively in varying $N$ logical zones. Under keen observation, one would notice that the $LUM1,\ldots,LUM4$ implemented with the proposed ISLS displays significant lower energy usage in comparison to $Base1,\ldots,Base4$ of Base installation. Also, the energy profiles of Fig.~\ref{profile1}, Fig.~\ref{profile2} and Fig.~\ref{profile4} proves that the ISLS utilizes only the luminaires in the occupied zones while other luminaires are conditionally auto-dimmed off at different levels. On the contrary, all the luminaires in the Base installation remain fully and uniformly active at all times once activated to ON state. To ensure accurate sensor reading at varying time of the day, this experiment did not exactly mimic regular real-world scenario where random mobility activities is often initiated by the occupant/s. However, in some instances, certain light-fixtures of the base installation were randomly turned off at certain intervals. Nevertheless, the percentage energy saving of the ISLS system in comparison to the base installation reached up to 80\% for single occupancy, 48\% for dual occupancy, and 26\% for quad-occupancy, respectively (refer to APPENDIX~\ref{Appendix1}).

This result proves the ISLS's autonomous ability to intelligently initiate state change of the connected luminaires (i.e., on--dim--off) with respect to daylight sensing and occupancy detection. To further validate this outcome, additional tests were conducted to ascertain the performance of the dimming profiles (i.e., the dimming curves) at difference preset illuminance values. %(--shown in Fig.~\ref{600lux} and Fig.~\ref{300lux}).

\begin{figure}[th]
\centering
\includegraphics[width=20pc]{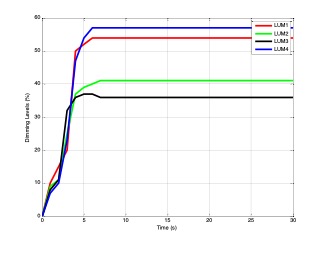}\\
\caption{Dimming curve of the ISLS when luminous flux is set to 600 Lux}\label{600lux}
\vspace{1mm}
\includegraphics[width=20pc]{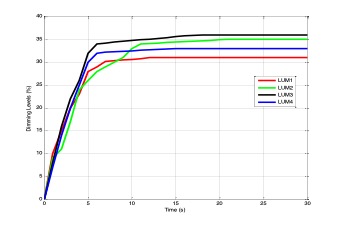}
\caption{Dimming curve of the ISLS when luminous flux is set to 300 Lux}\label{300lux}
\end{figure}

The dimming curve is the rate of change of light output as a function of the dimmer controller. In this task, a monotonic dimming trajectory is expected, i.e., the light level should not increase at any point along the curve when the dimmer is traveling down, vice versa. In this regard, we considered the trajectories for two different scenarios. The outputs of the obtained dimming trajectories when the luminous flux is set to 600 Lux and 300 Lux are shown in Fig. \ref{600lux} and Fig. \ref{300lux}, respectively. From observation (subject to occupant/s activities), both instances are monotonic. A monotonic trajectory means that as the dimmer level decreases (or increases), the light output consistently decreases (or increases) without any reversals. This consistency is critical for user predictability and ensuring that the system behaves as expected. According to \cite{Rea'00}, there is no limitation imposed by the detailed shape of the profile as long as the dimming profile is monotonic and meets dead travel range. Dead travel refers to the phenomenon where there is no actual or perceived change in the light output from a dimmable lamp, after there is a change in the dimming level.

\begin{figure*}%[ht!]
        \centering
            \subfigure[Daylight distribution when the luminaires are OFF.]
            {
                %\label{Time series plot of $lambda1$}
                \includegraphics[width=.42\textwidth]{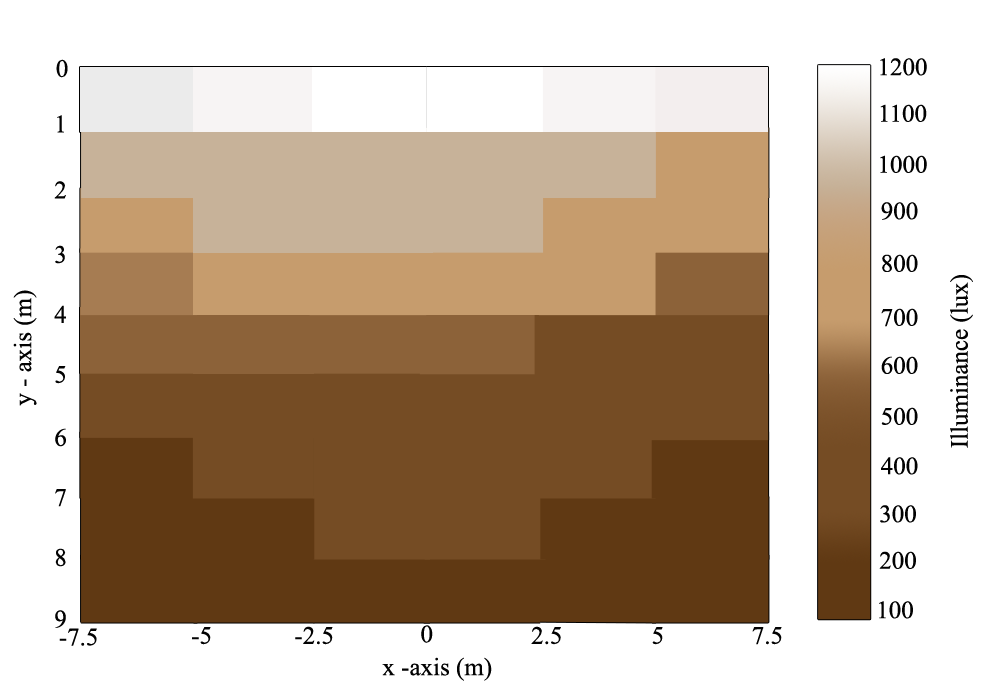} % .png .jpg ... according to supported graphics files
            }
            \subfigure[Early daytime luminaire Lux output with one occupant]
            {
                %\label{Time series plot of $lambda2$}
                \includegraphics[width=.46\textwidth]{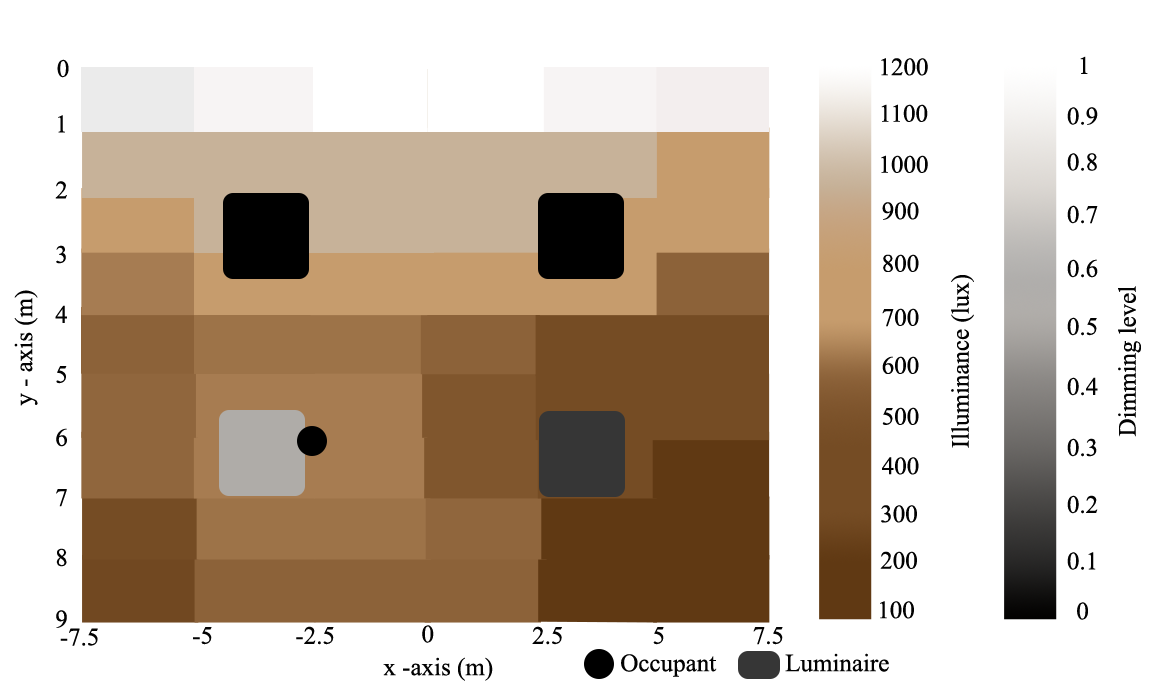} % .png .jpg ... according to supported graphics files
            }\\
            \subfigure[Peak daytime luminaire Lux output with two occupants]
            {
                %\label{Time series plot of $lambda3$}
                \includegraphics[width=.45\textwidth]{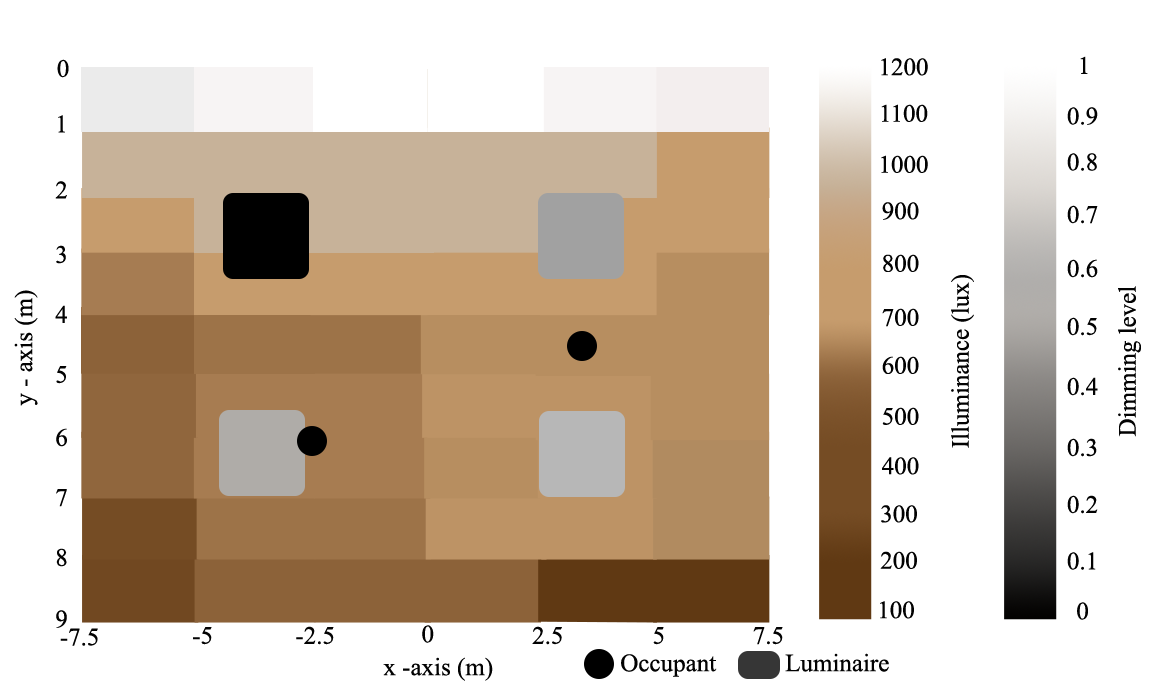} % .png .jpg ... according to supported graphics files
            } % for new row or line of subfigures
            \subfigure[Evening time luminaire Lux output with four occupants]
            {
                %\label{Time series plot of $lambda4$}
                \includegraphics[width=.45\textwidth]{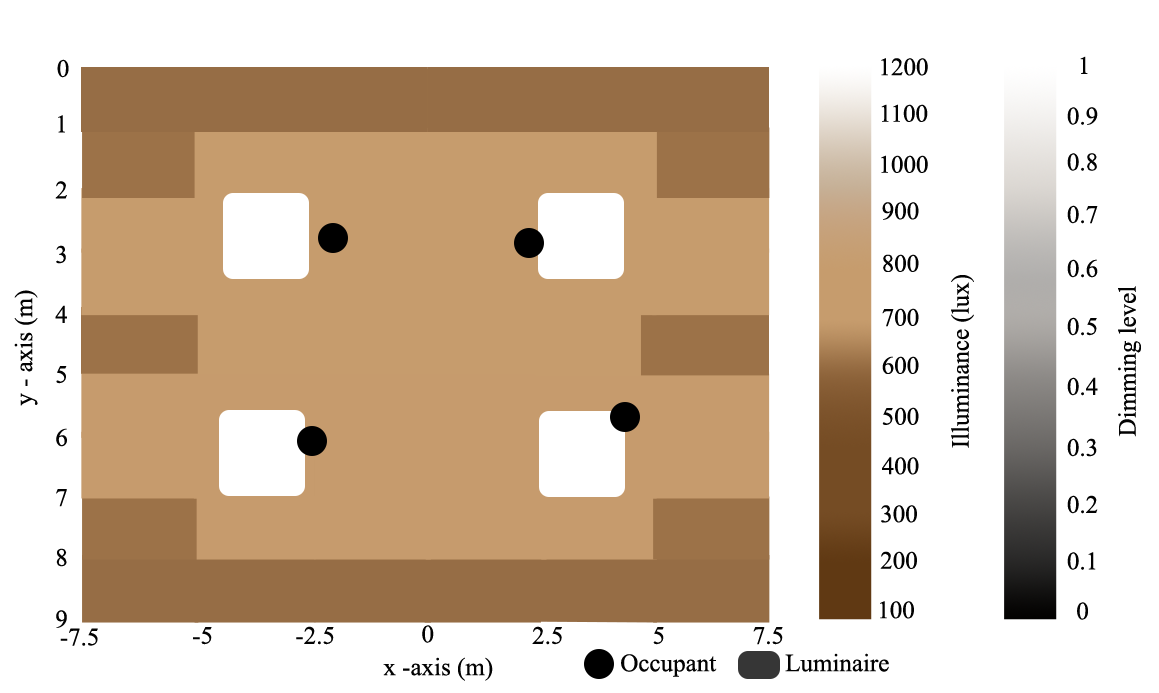} % .png .jpg ... according to supported graphics files
            }

        \caption{Light distribution across different time series in an ISLS-enabled workspace with varying occupancy levels}
        \label{zones}
        \end{figure*}

Finally, in order to substantiate the effectiveness of the real-world prototype experiment and verify the functional performance of the ISLS, we simulated sensor data under both daytime (work hours) and evening time (close of work hours) conditions in the absence of natural light sources using MATLAB. The simulation outcomes, depicted in Fig. \ref{zones}, demonstrate the system's capacity to adjust lighting in response to varying ambient conditions and occupancy levels. For example, under peak daylight conditions, as shown in Fig. \ref{zones}(a), the sensors correctly dimmed the luminaires to an off state, while Fig. \ref{zones}(b) illustrates an appropriate reduction in luminance in a workspace with one occupant during early morning hours. The obtained sensor data and achieved energy efficiency calculation during the occupancy scenarios shown in Fig. \ref{zones}(a)–(d) for one, two, and four user-activities, are provided in APPENDIX~\ref{Appendix1}. The outcome reinforces that the ISLS consistently modulates light output across different conditions —from a full daylight distribution with no occupancy, to early daytime, to peak daytime settings with one and two occupants respectively, and lastly, to a dusk scenario with up to four occupants. These results convincingly demonstrate that the ISLS is effective in adapting to both ambient lighting regardless of occupancy variations, for ensuring optimal performance in energy efficiency and user/s visual comfort.

\section{CONCLUSION}

In this paper, we presented a functional low-cost DIY prototype design and development of an IoT-based intelligent LED lighting system dubbed ISLS, featuring an illuminance control algorithm, and formulated through linear programming. The proposed ISLS system dynamically adjusts indoor lighting levels by integrating daylight and occupancy detection systematically for ensuring energy-efficient optimal illumination across various zones (i.e., indoor areas) of a building. We validated the effectiveness of the system through both simulated and real-world experiments to confirm its performance abilities under varying stress-tests; towards achieving minimal power consumption while delivering the required lighting levels that meets the control objectives of user/s photometric needs. The ISLS solution demonstrates scalability and adaptability, making it suitable for integration into existing traditional lighting infrastructures. Its noteworthy low cost and compliance with industry standards for dimming profiles and illumination patterns further enhances its appeal as a potential and viable energy conservation technology applicable to both commercial and residential settings.

%\section{ACKNOWLEDGMENTS}
%\noindent This work was supported in part by the National Science and Technology Council, Taiwan, under Grant MOST-110-2221-E-182-041-MY3.

\def\refname{REFERENCES}

\bibliographystyle{IEEEtran}
\bibliography{Lite-Ref}

\appendices

%\lipsum[1-2]%

\onecolumn {\section{Energy Savings Calculation\vspace{1.3\baselineskip}}\label{Appendix1}} 

\noindent To analyze the ISLS system performance for varying number of occupants (herein, $1,\cdots,4$) within a 12-hour activity in a workspace during daytime, we refer to the user/s scenario depicted in Fig.~\ref{zones}. 
\vspace{1.3\baselineskip}
\subsection{One occupant}
The data values generated by the sensors for a one-user occupant experiment is given in Table~\ref{tab:one_user} below.

\begin{table}[h]
    \caption{Energy Readings for Workspace with a Single Occupant}
    \label{tab:one_user}
    \centering
    \scriptsize  % To Reduce font size (for IEEE format)%
    \resizebox{\columnwidth}{!}{  % Ensures table fits within a single column
    \begin{tabular}{lccccccccc}
        \toprule
        \textbf{Hours} & \textbf{LUM1} & \textbf{LUM2} & \textbf{LUM3} & \textbf{LUM4} & \textbf{Base1} & \textbf{Base2} & \textbf{Base3} & \textbf{Base4} & \textbf{Energy Saving} \\
        \midrule
        6:00:00  & 0.0012 & 0.0012 & 0.0012 & 0.0012 & 0.012 & 0.012 & 0.012 & 0.012 & 0.012 \\
        7:00:00  & 0.012  & 0.012  & 0.012  & 0.012  & 0.012 & 0.012 & 0.012 & 0.012 & 0.012 \\
        8:00:00  & 0.006  & 0.009  & 0.009  & 0.009  & 0.012 & 0.012 & 0.012 & 0.012 & 0.012 \\
        9:00:00  & 0.003  & 0      & 0.003  & 0      & 0.012 & 0.012 & 0.012 & 0.012 & 0.018 \\
        10:00:00 & 0.003  & 0      & 0      & 0      & 0.012 & 0.012 & 0.012 & 0.012 & 0.012 \\
        11:00:00 & 0.003  & 0      & 0      & 0      & 0.012 & 0.012 & 0.012 & 0.012 & 0.012 \\
        12:00:00 & 0.003  & 0      & 0      & 0      & 0.012 & 0.012 & 0.012 & 0.012 & 0.012 \\
        13:00:00 & 0.003  & 0      & 0      & 0      & 0.012 & 0.012 & 0.012 & 0.012 & 0.012 \\
        14:00:00 & 0.003  & 0      & 0      & 0      & 0.012 & 0.012 & 0.012 & 0.012 & 0.012 \\
        15:00:00 & 0.003  & 0      & 0      & 0      & 0.012 & 0.012 & 0.012 & 0.012 & 0.012 \\
        16:00:00 & 0.003  & 0      & 0      & 0      & 0.012 & 0.012 & 0.012 & 0.012 & 0.012 \\
        17:00:00 & 0.003  & 0      & 0      & 0      & 0.012 & 0.012 & 0.012 & 0.012 & 0.012 \\
        \midrule
        \textbf{Total (KWhr)} & 0.0462 & 0.0222 & 0.0252 & 0.0222 & 0.144 & 0.144 & 0.144 & 0.144 & 0.15 \\
        \bottomrule
    \end{tabular}
    }
\end{table}

Let's presume that the user predominantly occupied the area covered by $LUM1$. With respect to sensor-obtained energy usage data values in Table~\ref{tab:one_user}, and owing to the single occupant's activities in the $LUM1$ logical zone depicted in Fig.~\ref{zones}(b), the calculation of the ISLS's total energy usage = 0.0462 + 0.0222 + 0.0252 + 0.0222 = 0.1158 KWhr. On the other hand, the total energy usage of the Base installation = 0.576 KWhr  (summed luminaries output), --given that they are not dimmable, the luminaires are thus consistently operating at full capacity in ON state. 

\noindent To comparatively juxtapose the performance of the ISLS over that of the Base installation, we calculate,
%Hence, to compara the ISLS versus the Base installation performance: 
\[
    \text{Total Energy Savings (KWhr)} = 0.576 - 0.1158 = 0.46 \text{ KWhr}
\]

\noindent And the percentage energy saved is given by,
\[
    \text{Percentage Energy Saved (\%)} = \left( \frac{100 \times 0.46}{0.576} \right) = 79.9\% \approx 80\%
\]

\vspace{1.3\baselineskip}
Similarly, the calculation of energy savings profile for two (2) and four (4)-occupant activity scenarios are provided below.  

%\vspace{1.3\baselineskip}
\subsection{Two occupants} 

Here, from Fig.~\ref{zones}(c), one user is positioned at coordinate (-2.5, 6) and the second user is located at (2.4, 4.5) --which is a zone overlapping two luminaires ($LUM2$ and $LUM4$). Hence, the presence sensors of both luminaires can partly detect occupant-2. The blend of the ISLS's illuminance output is such that a uniform illumination is achieved in the occupant's location while simultaneously accounting for day lighting effect. The sensors' readings are listed in Table~\ref{tab:two_users}. 

\begin{table}[h]
    \caption{Energy Readings for Workspace with Two Occupants}
    \label{tab:two_users}
    \centering
    \scriptsize  % To Reduce font size (for IEEE format)%
    \resizebox{\columnwidth}{!}{  % Ensures table fits within a single column
    \begin{tabular}{lccccccccc}
        \toprule
        \textbf{Hours} & \textbf{LUM1} & \textbf{LUM2} & \textbf{LUM3} & \textbf{LUM4} & \textbf{Base1} & \textbf{Base2} & \textbf{Base3} & \textbf{Base4} & \textbf{Energy Saving} \\
        \midrule
        6:00:00  & 0.0012 & 0.0012 & 0.0012 & 0.0012 & 0.012 & 0.012 & 0.012 & 0.012 & 0.012 \\
        7:00:00  & 0.012  & 0.012  & 0.012  & 0.012  & 0.012 & 0.012 & 0.012 & 0.012 & 0.012 \\
        8:00:00  & 0.012  & 0.006  & 0.003  & 0.006  & 0.012 & 0.012 & 0.012 & 0.012 & 0.012 \\
        9:00:00  & 0.006  & 0.003  & 0.012  & 0.012  & 0.012 & 0.012 & 0.012 & 0.012 & 0.018 \\
        10:00:00 & 0.006  & 0.003  & 0      & 0.012  & 0.012 & 0.012 & 0.012 & 0.012 & 0.012 \\
        11:00:00 & 0.006  & 0.003  & 0      & 0.012  & 0.012 & 0.012 & 0.012 & 0.012 & 0.012 \\
        12:00:00 & 0.006  & 0.003  & 0      & 0      & 0.012 & 0.012 & 0.012 & 0.012 & 0.012 \\
        13:00:00 & 0.006  & 0.003  & 0      & 0      & 0.012 & 0.012 & 0.012 & 0.012 & 0.012 \\
        14:00:00 & 0.006  & 0.003  & 0      & 0      & 0.012 & 0.012 & 0.012 & 0.012 & 0.012 \\
        15:00:00 & 0.006  & 0.006  & 0      & 0.012  & 0.012 & 0.012 & 0.012 & 0.012 & 0.012 \\
        16:00:00 & 0.012  & 0.012  & 0.012  & 0.012  & 0.012 & 0.012 & 0.012 & 0.012 & 0.012 \\
        17:00:00 & 0.012  & 0.012  & 0.012  & 0.012  & 0.012 & 0.012 & 0.012 & 0.012 & 0.012 \\
        \midrule
        \textbf{Total (KWhr)} & 0.0912 & 0.0672 & 0.0522 & 0.0912 & 0.144 & 0.144 & 0.144 & 0.144 & 0.15 \\
        \bottomrule
    \end{tabular}
    }
\end{table}

From the recorded values (in Table~\ref{tab:two_users}), the total energy consumed by the system during two-user occupancy is given by:

\[
    \text{Total Energy (KWhr)} = 0.0912 + 0.0672 + 0.0522 + 0.0912 = 0.30 \text{ KWhr}
\]

\noindent Recall that light fixtures from the Base installation are non-dimmable at ON state, therefore its total energy consumption per luminaire = 0.576 KWhr.

\noindent So, the total energy saved for two-occupant scenario is:

\[
    \text{Total Energy Saved (KWhr)} = 0.576 - 0.30 = 0.2759 \text{ KWhr}
\]

\noindent And calculating the percentage energy saved, we have,

\[
    \text{Percentage Energy Saved (\%)} = \left( \frac{100 \times 0.2759}{0.576} \right) = 47.9\% \approx 48\%
\]

%\vspace{1.3\baselineskip}
\subsection{Four occupants}

For the four-user scenario depicted in Fig.~\ref{zones}(d), it was observed that all the ISLS luminaires were turned ON and almost operating at maximum power, similarly to the Base installation. The obtained sensor readings are provided hereunder in Table~\ref{tab:four_users}. 

\begin{table}[h]
    \caption{Energy Readings for Workspace with Four Occupants}
    \label{tab:four_users}
    \centering
    \scriptsize  % To Reduce font size (for IEEE format)% 
    \resizebox{\columnwidth}{!}{  % Ensures table fits within a single column
    \begin{tabular}{lccccccccc}
        \toprule
        \textbf{Hours} & \textbf{LUM1} & \textbf{LUM2} & \textbf{LUM3} & \textbf{LUM4} & \textbf{Base1} & \textbf{Base2} & \textbf{Base3} & \textbf{Base4} & \textbf{Energy Saving} \\
        \midrule
        6:00:00  & 0.0012 & 0.0012 & 0.0012 & 0.0012 & 0.012 & 0.012 & 0.012 & 0.012 & 0.012 \\
        7:00:00  & 0.012  & 0.012  & 0.012  & 0.012  & 0.012 & 0.012 & 0.012 & 0.012 & 0.012 \\
        8:00:00  & 0.012  & 0.006  & 0.003  & 0.006  & 0.012 & 0.012 & 0.012 & 0.012 & 0.012 \\
        9:00:00  & 0.006  & 0.003  & 0.012  & 0.012  & 0.012 & 0.012 & 0.012 & 0.012 & 0.018 \\
        10:00:00 & 0.006  & 0.003  & 0      & 0.012  & 0.012 & 0.012 & 0.012 & 0.012 & 0.012 \\
        11:00:00 & 0.006  & 0.003  & 0      & 0.012  & 0.012 & 0.012 & 0.012 & 0.012 & 0.012 \\
        12:00:00 & 0.006  & 0.003  & 0      & 0      & 0.012 & 0.012 & 0.012 & 0.012 & 0.012 \\
        13:00:00 & 0.006  & 0.003  & 0      & 0      & 0.012 & 0.012 & 0.012 & 0.012 & 0.012 \\
        14:00:00 & 0.006  & 0.003  & 0      & 0      & 0.012 & 0.012 & 0.012 & 0.012 & 0.012 \\
        15:00:00 & 0.006  & 0.006  & 0      & 0.012  & 0.012 & 0.012 & 0.012 & 0.012 & 0.012 \\
        16:00:00 & 0.012  & 0.012  & 0.012  & 0.012  & 0.012 & 0.012 & 0.012 & 0.012 & 0.012 \\
        17:00:00 & 0.012  & 0.012  & 0.012  & 0.012  & 0.012 & 0.012 & 0.012 & 0.012 & 0.012 \\
        \midrule
        \textbf{Total (KWhr)} & 0.0912 & 0.0672 & 0.0522 & 0.0912 & 0.144 & 0.144 & 0.144 & 0.144 & 0.15 \\
        \bottomrule
    \end{tabular}
    }
\end{table}

\noindent From Table~\ref{tab:four_users}, the calculation for the total energy usage of ISLS per luminaire in the workspace logical zones is thus:

\[
    \text{Total Energy (KWhr)} = 0.105 + 0.105 + 0.108 + 0.108 = 0.426 \text{ KWhr}
\]

\noindent Calculating the total energy savings of the ISLS over the Base installation's $0.576$ KWhr summed power usage, we have,

\[
    \text{Total Energy Saved (KWhr)} = 0.576 - 0.426 = 0.15 \text{ KWhr}
\]

\noindent Therefore, the percentage energy saved during four occupants' activity is given by,

\[
    \text{Percentage Energy Saved (\%)} = \left( \frac{100 \times 0.15}{0.576} \right) = 26.04\% \approx 26\%
\]

\end{document}